\documentstyle[aps,prb,epsf]{revtex}
\draft
\begin{document}

\title{Room temperature Kondo effect in atom-surface scattering: 
dynamical $1/N$ approach}
\author{J. Merino and J. B. Marston}
\address{Department of Physics, Brown University, Providence, RI 02912-1843}
\date{April 9, 1998}  
\bigskip
\maketitle

\begin{abstract}
The Kondo effect may be observable in some atom-surface scattering 
experiments, in particular, those involving alkaline-earth atoms. 
By combining Keldysh techniques with the NCA approximation
to solve the time-dependent Newns-Anderson Hamiltonian in the
$U \rightarrow \infty$ limit, Shao, Nordlander, and Langreth 
found an anomalously strong  
surface-temperature dependence of the outgoing charge state fractions. 
Here we employ a dynamical $1/N$ expansion with a finite Coulomb interaction $U$
to give a more realistic description of the scattering process.
We test the accuracy of the $1/N$ expansion in the
$N = 1$ case of spinless fermions against the exact independent particle
solution.  We then compare results obtained in the $U \rightarrow \infty$ limit
with the NCA approximation at $N = 4$ and recover qualitative features already 
found.  Finally, we analyze the realistic situation of 
Ca atoms with $U = 5.8$ eV scattered off Cu(001) surfaces.  
Although the presence 
of the doubly-ionized Ca$^{++}$ species can change the absolute scattered 
Ca$^+$ yields, the temperature dependence is qualitatively
the same as that found in the 
$U \rightarrow \infty$ limit.  One of the main difficulties that 
experimentalists face in attempting to detect this effect is that 
the atomic velocity must be kept small enough to limit
kinematic smearing of the Fermi surface of the metal.

\end{abstract}

\bigskip
\section{Introduction}
\label{sec:Intro}

The interaction of a localized spin impurity with the conduction
electrons of a metal is a well known and interesting problem\cite{Hewson}.
The Kondo effect arises as a result of the interaction, and  
thermodynamic properties such as susceptibility and specific heat
have been thoroughly studied for many years, 
both theoretically and experimentally\cite{Kondo:69,Wilson:75,Heeger:69}.
Non-equilibrium properties of these systems, however, are not well understood.

Recent advances in the construction of small 
quantum dots permit a high degree of control in the coupling
between localized and extended electronic states.  Measurements of transport
through dots give clear evidence for the Kondo effect.\cite{Goldhaber:98}
Shao, Nordlander, and Langreth (SNL) recently pointed out\cite{Shao:96} that
atomic scattering off metal surfaces is another arena for the Kondo effect. 
The rich variety of atomic levels, workfunctions,
and couplings permits the investigation of many regimes. 
For example, as it encounters the surface, an
atom can evolve from the empty orbital regime into the local moment regime,
passing through the mixed valence state at intermediate distances from 
the surface.  
Strikingly, the Kondo scale can easily attain room temperature because the 
atomic level crosses the Fermi energy.
The velocity of the incoming ion may also be tuned to probe the 
different regimes.  However, 
only at sufficiently small velocities is there enough time for a well-developed
Kondo resonance to form.  The Kondo screening cloud manifests itself in 
enhanced neutralization probabilities at low speeds and reduced temperatures.
 
One theoretical approach used to analyze highly non-equilibrium
atom-surface systems is the Keldysh
Green's function technique\cite{Keldysh:69,Shao:94}, 
combined with the Non-Crossing Approximation (NCA)\cite{Shao:95,Wingreen:94}.
This approach typically takes the repulsive Coulomb interaction strength $U$
between electrons residing on the impurity or 
scattered atom to be infinite.  
Although NCA has been extended to treat the case of 
finite $U$, at least in the static case\cite{Schiller:93},
it can only be easily applied to the dynamical problem in the 
$U \rightarrow \infty$ limit.  However, renormalization group (RG) 
calculations show that the parameters of the system in the finite-$U$ 
case\cite{Haldane,Krishna-murthy:80} can differ significantly 
from those in the $U \rightarrow \infty$ limit; in particular the 
Kondo temperature depends on $U$.  Thus physically reasonable values for $U$
should be used in realistic calculations.  Finally, NCA has additional 
problems as it neglects vertex corrections. 
Fermi liquid relations are not recovered at zero temperature. 
Instead spurious behavior shows 
up in the spectral density of the impurity\cite{Costi:96}.

In this article we present an alternative approach, namely,
the dynamical $1/N$ method.  Details of the method have been
presented in [I]\cite{Marston:96}. Both resonant and Auger processes can be
treated in the same framework, as can finite $U$.  The method
has been used to study charge transfer between alkali ions, such as Li and Na,
and metal surfaces.  Semiquantitative predictions of Li(2p) excited state 
formation and final charge fractions agree well with
experiment\cite{Marston:93,Marston:96,Dahl}, in contrast with independent
particle theories which ignore the essential electron-electron correlations.
Here we extend the previous work of [I] by including 
a new sector in the wavefunction which describes the amplitude for
two particle-hole excitations in the metal.  As this sector is of order
$1/N^2$ the accuracy of the expansion is improved.  A further extension of the
solution permits treatment of non-zero substrate temperature.

This paper is organized as follows: In Sec. \ref{sec:theory} we briefly discuss
the theory and its new features.  In Sec. \ref{sec:results} we present
results in both the $U \rightarrow \infty$ limit and in the physical 
finite-U case. The $U \rightarrow \infty$ limit is studied for the purpose of 
comparing our $1/N$ results at $N = 4$ with previously published
NCA results\cite{Shao:96}.  As a further test of the $1/N$ expansion we 
also present results extended down to the spinless $N = 1$ limit.
In this case, there are exact independent particle results which can be  
used to test the accuracy of the $1/N$ expansion.  Finally, the physically 
relevant case of finite-$U$ and $N = 2$ is examined to determine whether or
not Kondo effects are accessible to charge transfer experiments. 
We conclude in section \ref{sec:conclusions} with a discussion of the 
conditions which must be satisfied in charge transfer 
experiments which search for the Kondo effect. 
  
\section{Theoretical background}
\label{sec:theory}
Independent-particle descriptions of charge transfer fail to account
for many interesting and non-trivial phenomena.  Here we briefly review the
main features of correlated model we employ and its approximate $1/N$ solution.
Details can be found in [I]\cite{Marston:96}.

\subsection{The model}
\label{subsec:model}
Our model for atom-surface system is the generalized time-dependent 
Newns-Anderson Hamiltonian.  For concreteness we first consider the case
of alkali atoms, and then later show how a particle-hole transformation 
permits the description of alkaline-earth atoms.  To organize the solution 
systematically in powers of $1/N$, the atomic orbital is taken
to be $N$-fold degenerate.  The physical case of electrons with
up and down spins can then be studied by setting $N = 2$. 
\begin{eqnarray}
H(t) &=& \sum_a [\epsilon^{(1)}_a(z) \hat P_1  + \epsilon^{(2)}_a(z) \hat P_2]~
\ c_a^{\dag \sigma} c_{a \sigma}
 +  \sum_k \epsilon_k\ c_k^{\dag \sigma} c_{k \sigma}
\nonumber \\
&+& N^{-1/2} \sum_{a;\ k} \{ [V^{(1)}_{a;k}(z) \hat P_1
+ V^{(2)}_{a;k}(z) \hat P_2]
\ c_a^{\dag \sigma} c_{k \sigma} + H.c \}
\nonumber \\
&+&  {{1}\over{2}} ~ \sum_a U_{aa} n_a (n_a - 1)
+ \sum_{a > b} U_{ab} n_a n_b ~.
\label{O.H}
\end{eqnarray}
where the indices $a$ labels the discrete set of atom states and $k$
labels the continuum of electron states in the metal.
Projection operators $\hat P_1$ and $\hat P_2$ project respectively onto the
one and two electron subspaces and can be written explicitly
in terms of the occupation number
operators.  Implicit summation over repeated raised and lowered spin
indices, $\sigma = 1, 2, \cdots, N$, is assumed.

Parameters $\epsilon_a^{(1)}(z)$, $\epsilon_a^{(2)}(z)$, $V_{a;k}^{(1)}(z)$ 
and $V_{a;k}^{(2)}(z)$ are, respectively, the orbital energies and matrix 
elements for 1 and 2 electrons.  The use of the projector operators
enable us to use different couplings depending on the
number of electrons in the atom.  The couplings are divided by
the square root of the degeneracy $\sqrt{N}$ so that the atomic width
$N \Gamma$ remains finite and well-defined in the $N \rightarrow \infty$ limit.
As excited states of the atom with two electrons
occupy orbitals high in energy they are excluded.
We therefore take $U_{ab} \rightarrow \infty$ when $a \neq 0$ and $b \neq 0$.
In the limit $U = U_{00} \rightarrow \infty$  
the above Hamiltonian reduces to a simpler, but still nontrivial, form
as now the atom can be occupied by at most one electron:
\begin{eqnarray}
H(t) &=& \sum_a \epsilon^{(1)}_a(z)~ \hat P_1~
c_a^{\dag \sigma} c_{a \sigma}
+ \sum_k \epsilon_k\ c_k^{\dag \sigma} c_{k \sigma}
\nonumber \\
&+& N^{-1/2} \sum_{a;\ k} \{ V^{(1)}_{a;k}(z)~ \hat P_1~
c_a^{\dag \sigma} c_{k \sigma} + H.c \}
\nonumber \\
\label{I.H}
\end{eqnarray}
Setting $U = 0$ and $N = 1$ in Eq. (\ref{O.H}), on the other hand,
yields the spinless independent-particle Newns-Anderson 
Hamiltonian\cite{Newns:69,Brako:81}.

Explicit time dependence enters through the classical ion trajectory.  As
the ion's kinetic energy is typically much larger than the electronic energy, 
the trajectory to a good approximation may be considered fixed.  Thus the
distance of the ion away from the surface may be written simply as: 
\begin{eqnarray}
z(t) &=& z_f - u_i~ t;\ \ \ t \leq t_{turn} \equiv (z_f - z_0)/u_i.
\nonumber \\
&=& z_0 + u_f~ (t - t_{turn});\ \ \ t > t_{turn}.
\label{trajectory}
\end{eqnarray}
In words, at time $t = 0$ an incoming atom at position $z = z_f$ moves 
at perpendicular velocity $u_i$ towards the surface.  At the turning point, 
$z = z_0$ it reverses course and leaves with velocity $u_f$.  We also 
consider trajectories which start from the surface at $t = 0$ and $z = z_0$.
Perpendicular incoming and outgoing velocities $u_i \geq u_f$ can differ
to account for loss of kinetic energy during the collision.  We assume in
the following that energy deposited as lattice motion is decoupled from 
the electronic degrees of freedom, at least over the relatively
short time scale of the atom-surface interaction.

The energy of a neutral alkali atom relative to the Fermi energy, 
which we denote by $E[A^0]$, shifts 
due to the image potential and saturates close to the surface: 
\begin{eqnarray}
E[A^0] &=& {\epsilon^{(1)}_a(z) } =  W - I_a +
(1/V_{maxI}^2 + 16 (z - z_{im})^2 / e^4)^{-1/2},\ \ \ z > z_{im}
\nonumber \\
&=& W - I_a + V_{maxI},\ \ \ z < z_{im}
\label{alkshift}
\end{eqnarray}
Here $I_a$ is the ionization energy required to remove one electron from
the a-orbital, $W$ is the workfunction 
and $V_{maxI}$ is the maximum image shift of the
atomic level as it approaches the surface.

It is important to notice that in the Hamiltonian of Eq. (\ref{O.H}), 
the unoccupied atomic
state, $A^{+}$, is defined to have zero energy so that it is $E[A^{0}]$ 
and not $E[A^{+}]$ which shifts.  Likewise the energy of the 
doubly-occupied configuration does not shift as the energy required to convert 
a negative alkali ion into a positive ion is constant
at all distances from the surface.  Thus the combination 
$2\epsilon^{(2)}_a(z) + U_{aa}$ is independent of $z(t)$ and we obtain simply:
\begin{equation}
\epsilon^{(2)}_a = W - I_a \ .
\label{affinshift}
\end{equation}
Finally the Coulomb interaction can be expressed as the difference 
between the affinity and the ionization levels: $U \equiv U_{00} = A + I_0$,
where the affinity is a negative quantity.

\subsection{The model for alkaline-earth atoms}
\label{subsec:alkearth}
While a free alkali atom, say Li, has one electron in the valence s-orbital, 
Li($2s^1$),
an alkaline-earth atom, say Ca, has a filled s-shell configuration, Ca($4s^2$).
Due to the image potential, the neutral Li atom tends to ionize by 
transfer of its valence electron close to the Cu surface, yielding the 
closed-shell configuration $Li^+(1s^2)$.   
In contrast, the most probable state of a Ca atom close to 
the surface is $Ca^+(4s^1)$ which has an unpaired spin in the valence orbital.  
This difference is what makes the alkaline-earth atoms interesting candidates
for revealing the properties of the Kondo screening cloud 
near metal surfaces.  We may easily treat the case of alkaline-earth
atoms (denoted by AE) within the wavefunction ansatz discussed below
in the next subsection by performing a 
particle-hole transformation.  This canonical transformation is implemented by 
switching all of the electron creation and annihilation operators: 
$c_a \rightarrow -c^+_a $,  $c^+_a \rightarrow -c_a$,
$c_k \rightarrow c^+_k$, and $c^+_k \rightarrow c_k$.  Minus signs have
been introduced to keep the $V_{a;k}$ matrix elements invariant.
If $\epsilon_k$ is particle-hole symmetric as we assume, the transformation 
leaves the Hamiltonian unchanged up to an irrelevant additive constant as
long as the energy of the atomic orbital is also reversed in sign.  Thus
$\epsilon^{(1)}_a$ should now be interpreted as the energy  
 time of one hole in the s-orbital of the atom, corresponding to a 
positive alkaline-earth ion, $AE^{+}$.  
In this case it is appropriate to set the energy of the neutral atom 
to zero, $E[AE^0] = 0$.  Also the image shift is reversed in sign.
Again taking into account image saturation close to the surface, the level 
variation as a function of distance is given by:
\begin{eqnarray}
E[AE^+] &=& {\epsilon^{(1)}_a(z) } =  I_a - W -
(1/V_{maxI}^2 + 16 (z - z_{im})^2 / e^4)^{-1/2},\ \ \ z > z_{im}
\nonumber \\
&=& I_a - W - V_{maxI},\ \ \ z < z_{im}.
\label{alkearthshift}
\end{eqnarray}
The total energy of the two-hole configuration, $AE^{++}$, then shifts like 
\begin{eqnarray}
E[AE^{++}] &=& {2\epsilon^{(2)}_a(z) } + U  =  2 (I_a - W) + U -
4(1/V_{maxII}^2 + 16 (z - z_{im})^2 / e^4)^{-1/2},\ \ \ z > z_{im}
\nonumber \\
&=& 2 (I_a - W) + U  - 4 V_{maxII},\ \ \ z < z_{im}.
\label{doubleshift}
\end{eqnarray}
Here $V_{maxI}$ and $V_{maxII}$ are respectively the maximum shifts in 
the one and two-hole orbital energies.  Far from the surface, the 
doubly-ionized level shifts four times as much as the singly ionized level. 
Consequently it can play an important role even when $U$ is rather large.

\subsection{Approximate solution of the model}

The time-dependent many-body wavefunction is systematically expanded
into sectors which contain an increasing number of particle-hole pairs
in the metal.  Amplitudes for sectors with more and more particle-hole
pairs are reduced by powers of $1/N$.  This approach was used with success
by Varma and Yafet\cite{Varma:76} and by Gunnarsson and 
Sch\"onhammer\cite{Gunnarsson:83,Gunnarsson:85}, 
to describe the static ground state properties 
and spectral densities of metal impurities such as Ce.  It was first applied
to the dynamical problem in 1985 by Brako and Newns\cite{Brako:85}; 
three sectors were included in this pioneering work.  Subsequent work 
work extended this ansatz\cite{Amos:92,Marston:93,Marston:96}.
The improved ansatz for our wavefunction now has six sectors (for details,
see [I]):
\begin{eqnarray}
| \Psi(t) \rangle &=& f(t) | 0 \rangle +
\sum_{a;\ k} b_{a;k}(t) |a; k \rangle
+ \sum_{k,\ L} e_{L,k}(t) |L, k \rangle
+ \sum_{q<k} d_{k,q}(t) |k, q \rangle
\nonumber \\
&+& \sum_{a;\ L,\ q<k} s_{a; L,k,q}(t) {|a; L, k, q \rangle}^{S}
+ \sum_{a;\ L,\ q<k} a_{a; L,k,q}(t) {|a; L, k, q \rangle}^{A}
\nonumber \\
&+& \underbrace{ \sum_{L>P,\ q<k} g_{L,P,k,q}(t) 
{|L, P, k, q \rangle}^{S}
+\sum_{L>P,\ q<k} h_{L,P,k,q}(t) {|L, P, k, q \rangle}^{A}}
_{new~ sector}
\nonumber \\
&+& \{ rest\ of\ Hilbert\ space \}\ .
\label{O.PSI}
\end{eqnarray}
Here $|0\rangle$ represents the unperturbed filled Fermi sea of the metal
and either an empty alkali valence orbital ($A^+$) or a filled alkaline-earth 
orbital ($AE^0$).  As the Hamiltonian conserves spin, the many-body 
wavefunction describes the spin singlet sector of the Hilbert space.   
The physical meaning of each of the amplitudes differs for alkali and 
alkaline earth atoms.  For alkalis
$e_{L,k}(t)$ is the amplitude for a positive ion with an empty valence
shell, one electron in state $L>k_F$ of the metal, and one hole in state 
$k<k_F$ of the metal where $k_F$ is the Fermi momentum.  
For alkaline-earths, on the other hand, $e_{L,k}(t)$ is the 
amplitude for a neutral atom with a filled s-orbital, an electron in state 
$k>k_F$ of the metal and a hole in state $L<k_F$ in the metal.

As the last sector included in the 
above expansion Eq. \ref{O.PSI} is $O(1/N^2)$, 
the accuracy of the $1/N$ expansion
has been improved compared to [I] which only considered terms up to $O(1/N)$.
The orthonormal basis for this two particle-hole pair sector can be written 
as the superposition of two amplitudes: 
\begin{eqnarray}
{|L, P, k, q \rangle}^{S}
&\equiv& {{1}\over{\sqrt{2N(N-1)}}}~ \lbrace\ c_{L}^{\dagger \alpha}
c_{k \alpha} c_{P}^{\dagger \beta} c_{q \beta} | 0 \rangle\
+ \ c_{L}^{\dagger \alpha} c_{q \alpha} c_{P}^{\dagger \beta}
c_{k \beta} | 0 \rangle\ \rbrace.
\nonumber\\
{|L,P, k, q \rangle}^{A}
&\equiv& {{1}\over{\sqrt{2N(N+1)}}}~ \lbrace\ c_{L}^{\dagger \alpha}
c_{k \alpha} c_{P}^{\dagger \beta} c_{q \beta} | 0 \rangle\
- \ c_{L}^{\dagger \alpha} c_{q \alpha} c_{P}^{\dagger \beta}
c_{k \beta} | 0 \rangle\ \rbrace.
\label{N.Sector}
\end{eqnarray}
The decomposition into symmetric and antisymmetric parts 
reflects the different ways that two electrons can
move from two occupied states of the metal labeled 
by $k$ and $q$ to two unoccupied states labeled by $P$ and $L$.
The sector is depicted, along with all the lower order sectors, in Fig. 8(b)
of [I].  The retention of the O$(1/N^2)$ sector improves 
loss of memory, the experimental observation that the outgoing charge 
probability distribution of alkali and halogen ions is independent of 
the incoming charge state\cite{Marston:96}, but otherwise does not alter 
any of the results presented below qualitatively.

The equations of motion are obtained by projecting the many-body
Schr\"odinger equation, 
$i {d\over{dt}} | \Psi(t) \rangle = \hat{H}(t) | \Psi(t) \rangle$, onto each 
sector of the retained Hilbert space\cite{Marston:96}.  
As a practical matter, the metal is modeled by a 
set of $M$ discrete levels both above and below the Fermi energy, and 
the Fermi energy is defined to be zero.  
Typically $M = 20$ in our calculations, but we
test the numerical accuracy by using larger values of $M$ up to $M = 50$.
The states in the metal are sampled unevenly: the mesh is made much finer
near the Fermi energy to account for particle-hole excitations
of low energy, less than $100$ K.  To be specific, the energy levels have
the following form:
\begin{equation}
\epsilon_k = \pm {{D}\over{e^\gamma - 1}}~ [e^{\gamma (k - 1/2) / M} - 1],\ \ \ 
k = 1, 2, \cdots, M\ .
\end{equation}
Here $D$ is the half-bandwidth of the metal.
Thus, the spacing of the energy levels close to the Fermi energy is 
reduced from the evenly spaced energy interval of $D/M$ by a factor of
$\gamma / (e^\gamma - 1) \approx 1/13$ for a typical choice of the sampling
parameter $\gamma = 4$.  The density of states $\rho \equiv M/D$, 
however, is kept uniform by including compensating weights
\begin{equation}
\sqrt{{{\gamma e^{\gamma (k - 1/2) / M}}\over{e^\gamma - 1}}}
\end{equation}
in all sums over the continuum of metal states $k$ which appear 
in the equations of motion for the component amplitudes. 
Numerical solution of the $O(M^4)$ 
coupled differential equations is performed with a 
fourth-order, Runge-Kutta algorithm with adaptive time steps.  
The double-precision code, written in C using structures to organize
the different sectors of the many-body wavefunction, runs in a 
vectorized form on Cray EL-98 and J-916 machines\cite{code}.   

Non-zero temperature is incorporated by sampling excited states, weighted by
the Boltzmann factor.  That is, at initial time $t = 0$ when the atom
is either far from the surface, or at the point of closest approach, 
the operator 
\begin{equation}
\exp \{-{{1}\over{2}}~ \beta~ \hat{H}(0)\} = \lim_{N \rightarrow \infty}
[1 - {{\beta}\over{2N}}~ \hat{H}(0)]^N 
\end{equation}
is applied to a set of random initial wavefunctions, and each resulting 
wavefunction is then integrated forward in time until the atom is
far from the surface.  Observables such as the atomic occupancy are then 
computed, and an average over the ensemble of initial states is 
performed, yielding the desired thermal average.  
In practice only 30 random states are required to yield accurate results.

\section{Results and discussion}
\label{sec:results}
   
We first compare the $1/N$ expansion with the NCA approach in the 
$U \rightarrow \infty$ limit.  Then we study the more 
realistic case of finite $U = 5.8$ eV which describes a scattered calcium atom. 

\subsection{$U \rightarrow \infty$ case}  
   
We begin by considering an alkaline-earth atom initially in equilibrium 
close to the surface.  At time $t = 0$ it leaves the surface at constant 
perpendicular velocity $u_f$, and we follow the charge transfer along the 
outgoing trajectory. 
For the purpose of comparing with previous work, the atom-surface system 
is modeled with the same
parameters as those employed by SNL\cite{Shao:96}.  
By setting $V_{max} \rightarrow \infty$ we obtain the pure image
shift of the level: $e_a(z) = e_a(\infty) - e^2 / 4 (z-z_{im})$
with the image plane positioned as it should be for a Cu surface: 
$z_{im}=1.0$ a.u.
away from the jellium edge.  The energy of the atomic level far from the 
surface is set to $e_a(\infty)= 1$ eV above the Fermi energy.
This positive level is the energy of one hole in a alkaline-earth atom,
as explained above in Sec. \ref{sec:theory}.  We set the turning
point in the trajectory of the ion at $z_0 = 3$ a.u. 
The half-width of the level, $\Delta_a$, is defined using the independent
particle formula as $\Delta_a$ = $\pi \rho V_{a}^2$,  
and the half-bandwidth of the metal is set to $D = 5$ eV.  The atomic widths
$\Delta$ can be obtained from first-principle independent-particle 
calculations.\cite{Merino:96,Nordlander:88,Kurpick:96,Gauyacq:93}
Following SNL\cite{Shao:95}, we assume a simple exponential
drop off for the width by setting:
$\Delta_a(z) = \Delta_0~ e^{-\alpha z }$ where $\Delta_0 = 0.75$ 
and $\alpha = 0.65$ a.u.

\begin{figure}
\center
\epsfxsize = 10cm \epsfbox{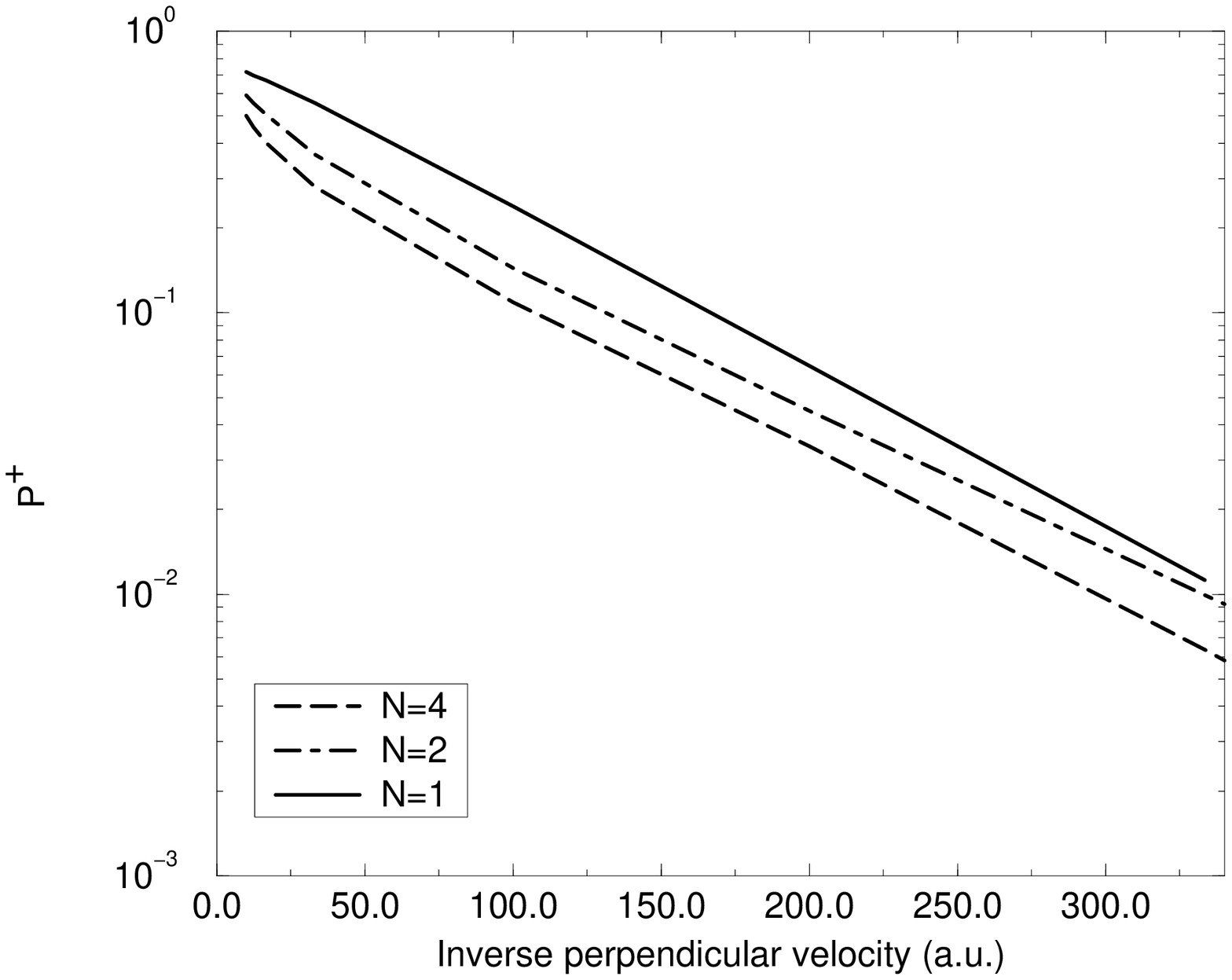}
\caption{
Final positive fractions at $T = 0$ K as a function of 
the inverse of the velocity for three different
degeneracies $N = 1$, 2, 4.  The atom starts from equilibrium at $z_0 = 3$ a.u.}
\label{fig1}
\end{figure}

In Fig. \ref{fig1} we plot the final positive populations versus the
inverse of the velocity for three
different values of the degeneracy factor $N = 1$, 2 and 4, 
and at zero surface temperature, $T = 0$ K. 
As the electron-hole transformation has been employed here to describe the
alkaline-earth atom, the $P^+$ charge fractions plotted in Fig. \ref{fig1} 
should be compared with the {\it neutral} occupancies plotted in Fig. 4 of 
SNL\cite{Shao:96}.  Fig. \ref{fig1} shows that the 
final positive fractions decrease with increasing
degeneracy factor $N$. This is as expected from static calculations.
For a fixed atomic position, the spectral weight right above the 
Fermi energy is enhanced as the degeneracy of the atomic level is increased, 
reducing the occupied fraction below the Fermi energy.  This effect can 
also be understood from the Friedel-Langreth sum rule.\cite{Langreth:66} 

Also apparent in the semi-log plot of Fig. \ref{fig1} is a breakdown
in linearity in $N = 2$ and $N = 4$ cases 
in contrast to the spinless $N = 1$ case. Although 
a similar upturn at high velocities is found within the NCA approach,
the effect here is quantitatively smaller.
SNL\cite{Shao:95} explain the upturn as a non-adiabatic 
effect due to the existence of two
time scales in the problem, a slow time scale set
by the Kondo temperature and a rapid time scale set by the level width. 
Another interesting feature in the figure is the
convergence of the $N = 2$ and $N = 4$ yields to the spinless
$N = 1$ yield at the lowest perpendicular velocities.
As the velocity is decreased the freezing
distance moves outwards and eventually the atom attains the empty orbital 
regime.  In this regime the many-body and the independent particle solutions
agree as the effect of the electron-electron interaction is negligible.

We would not expect the $1/N$ expansion to give accurate results at
$N = 1$.  Yet in the static case it has been demonstrated\cite{Gunnarsson:83} 
that the $1/N$ expansion is surprisingly accurate even down to $N = 1$. 
In contrast, the NCA approach retains spurious remnants of the Kondo 
peak in the spectral density of the impurity at $N = 1$.
The breakdown in NCA, an approximate solution which sums up a selected set 
of $O(1/N)$ diagrams, can be attributed to the neglect of 
vertex corrections\cite{Bickers:87} which are one order higher, O($1/N^2$).  
The $1/N$ expansion includes these diagrams, but does not resum them.  
As a consequence, static calculations of the impurity spectral density 
within the $1/N$ expansion can only yield a delta function peak at the Kondo 
resonance.  The integrated weight is correct, but 
only within a resummation scheme like NCA does it broaden out with non-zero 
width\cite{Bickers:87}.

In Table \ref{table1} we present values for the final charge fractions obtained
from the exact independent particle solution to the spinless 
Hamiltonian\cite{Marston:96}
and the $1/N$ approximation at $N = 1$ and $T = 0$ K as a function of 
different perpendicular velocities.
\begin{table}
\caption{Comparison at different perpendicular velocities of the exact 
independent particle solution to the spinless
Hamiltonian and the $1/N$ expansion at $N = 1$.} 
\vskip0.1in
\label{table1}
\begin{tabular}{dddd}
$u_f$ (a.u.) & $1/u_f$ (a.u.)& $P^+$(exact) & $P^+~ (1/N)$ \\
\hline
0.1  & 10.0 & 0.717 & 0.630 \\
0.03  & 33.3 & 0.555 & 0.417 \\
0.01  & 100.0 & 0.2384 & 0.189\\
0.003 & 333.3 & 0.0112 & 0.0143\\
\end{tabular}
\end{table}
The number of particle-hole pairs formed is always small, typically less
than one\cite{Marston:96}, and this fact 
accounts for why sectors in the wavefunction of order higher
than O($1/N^2$) do not contribute significantly to the final fractions. 
Thus the $1/N$ expansion retains high accuracy even in the most unfavorable
case of $N = 1$. 

\begin{figure}
\center
\epsfxsize = 10cm \epsfbox{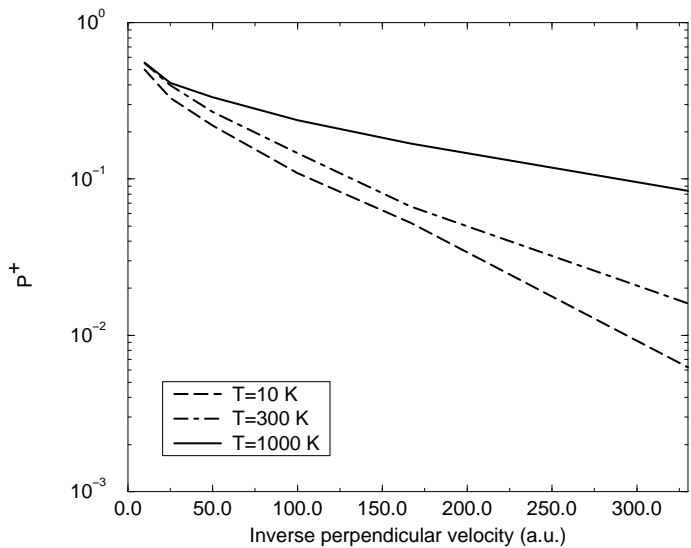}
\caption{
\label{fig2}
Final positive fractions versus the inverse of the velocity
for four different temperatures: $T = 10$, 300, 1000 K in 
the case $N = 4$ and $U \rightarrow \infty$.  The atom starts from 
equilibrium at $z_0 = 3$ a.u.}
\end{figure}

Up till now we have only considered the case of a zero-temperature metal   
surface.  We now study the temperature dependence of the final scattered 
fractions.  Although spectral densities are of interest and        
contain more information, we restrict ourselves to the calculation of 
physical observables: the occupancies of the various sectors along the 
trajectory.  These quantities give indirect information of the
highly correlated states formed close to the surface.
In Fig. \ref{fig2}, we present the final fractions versus
the inverse of the velocity for three different temperatures 
$T = 10$ K, $300$ K and $1000$ K and with a spin degeneracy
$N=4$.  We observe that the final positive charge 
fractions are enhanced as the temperature is increased.
The $1/N$ method reproduces the qualitative temperature dependence
found in the NCA approach (compare with Fig. 4 of SNL\cite{Shao:96}). 
We do not expect to get the same absolute yields 
as the NCA.  For example, while we have used a constant density of states 
to model the substrate, SNL\cite{Shao:96} use instead
a semielliptical band.  Because the Kondo scale depends on both
high and low energy processes, its numerical value will differ depending
on the choice of the band structure.  Nevertheless,  
the good qualitative and semiquantitative agreement between the 
two approaches suggests that the NCA breakdown at low temperatures is not
a problem here.\cite{Langreth:private} 

\begin{figure}
\center
\epsfxsize = 10cm \epsfbox{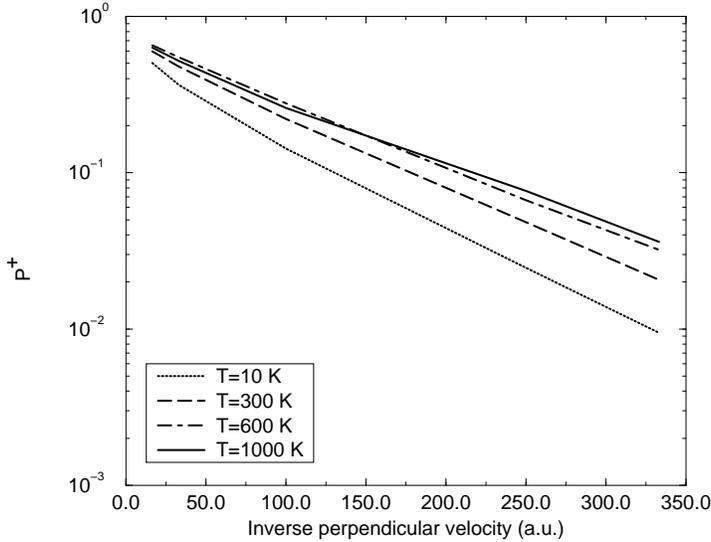}
\caption{
\label{fig3}
The same as in Fig. 2 but setting $N = 2$.}
\end{figure}

\begin{figure}
\center
\epsfxsize = 10cm \epsfbox{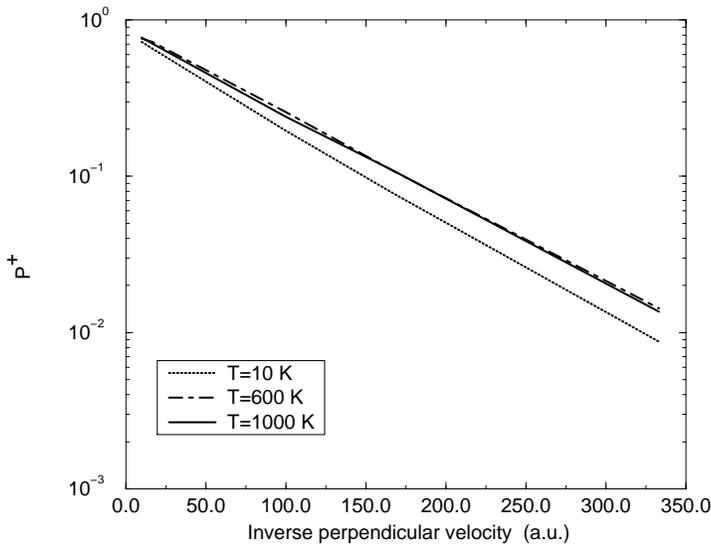}
\caption{
\label{fig4}
The same as in Fig. 2 but setting $N = 1$.}
\end{figure}

As emphasized by SNL\cite{Shao:96}
the final fractions show strong temperature dependence 
which cannot be understood from an independent particle 
picture.  Again setting $N = 1$ the results of Fig. \ref{fig4} 
show that the temperature dependence is in fact negligible as expected. 
There is some small temperature dependence at high temperatures, but this 
is just the usual thermal effect encountered previously in simple
independent particle pictures\cite{Los:90}.
  
\begin{figure}
\center
\epsfxsize = 10cm \epsfbox{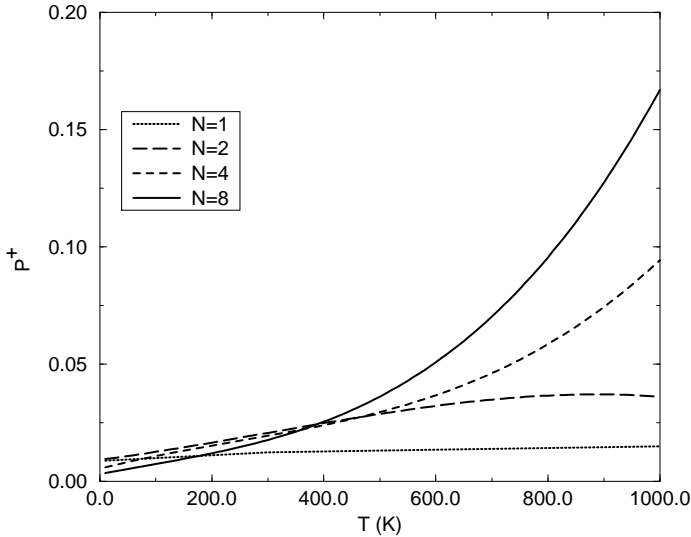}
\caption{
\label{fig5}
Final positive fractions versus temperature for a fixed velocity
of $u_f = 0.003$ a.u. and different
degeneracies of the atom: $N = 1$, 2, 4 and 8.  The $U \rightarrow \infty$ 
limit has been taken, and the atom starts at $z_0 = 3$ a.u. in
the equilibrium ground state.}
\end{figure}

In Fig. \ref{fig5} we have also plotted the temperature dependence of the 
final fractions  for four different degeneracies of the impurity: 
$N = 1$, $2$, $4$ and $8$ and at a fixed velocity of $u_f = 0.003$ a.u. 
While in the $N = 1$ spinless case the final 
positive fractions remain fairly constant, 
for $N > 1$ there is a strong temperature dependence of the final fractions 
with the temperature.  As expected the temperature dependence is 
stronger for larger degeneracies of the atomic orbital.
Thus, as the temperature of the substrate decreases, neutralization
takes place more efficiently.  The effect may be understood from the  
static picture, as the screening of the unpaired spin of an impurity
in a metal becomes more effective as the temperature drops below the 
Kondo scale.  In the dynamical problem the final outgoing charge fractions 
reflect this fact as screening of both the unpaired spin and charge occur
together.

Insight into the anomalous temperature dependence 
can be obtained by analyzing the renormalization of the parameters of the 
system when the atom is held at a fixed distance from the surface,
at equilibrium.  Table \ref{table2} records the values of  
the bare atomic level, the unrenormalized width, and 
Haldane's scaling invariant\cite{Haldane}, $E^*$, for different distances 
at zero temperature. Although perturbative scaling can only be safely applied 
in the weak coupling regime $\Delta_a \ll D$ and $U \gg D \gg |\epsilon_a|$,  
at sufficiently large distances from the surface (see Table) these conditions 
hold.  The fraction $|E^*/\Delta_a|$ determines what regime the impurity is in. 
For most of the trajectory, the impurity is in the mixed-valence
regime ($|E^*/\Delta_a| < 1$).  In this regime, $E^{*}$ can be identified
as the renormalized level\cite{Hewson} and we see from Table \ref{table2}
that the level is just above the Fermi energy in the 
relevant spatial region where the level width is 
sufficiently large to permit significant charge transfer.
As the temperature is increased
the renormalized level is thermally populated with holes from 
the metal enhancing the final positive fractions.
For distances larger than $z = 8.0$ a.u, the system reaches the
empty orbital regime and the bare and renormalized energy levels 
converge as many-body effects are negligible.  Of course 
in the spinless $N = 1$ case, the level remains unrenormalized, and 
there is negligible temperature dependence 
in the final fractions\cite{Brunner:97}.

\begin{table}
\caption{Haldane's scaling invariant and renormalized levels for $N = 2$. 
All energies are given in eV.} 
\vskip0.1in

\label{table2}
\begin{tabular}{ddddd}
$z$ (a.u.) & $\epsilon_a(z)$ & $\Delta_a(z)$ & $E^*(z)$  
& $E^*/\Delta_a$ \\
\hline
3.0  & -2.4 & 2.90 & -1.482 &-0.510  \\
4.0  & -1.27 & 1.516 & -0.474 &-0.312  \\
5.0  & -0.7 & 0.7913 & -0.122 & -0.155 \\
6.0  & -0.36& 0.4131 & 0.0267 & 0.0647 \\
7.0  & -0.13 & 0.2156 & 0.113 & 0.523 \\
8.0  & 0.028 & 0.1126 & 0.180 & 1.602 \\
9.0  & 0.150 & 0.0587 & 0.2412 & 4.104 \\
10.0 & 0.244 & 0.03  &  0.30 & 9.722 \\
\end{tabular}
\end{table}

When the atom is sufficiently close to the surface it enters the
local moment regime.  If the atom stays for a 
long time in this situation, the electrons at the Fermi level
couple antiferromagnetically with the unpaired spin of the
atom and screen the impurity spin.  An estimate of the Kondo temperature using
the Bethe-ansatz\cite{Hewson} solution in the $U \rightarrow \infty$ 
at $z = 3$ a.u. gives $k_B T_K \approx 0.1$ eV which approximately agrees with 
the position of the Kondo peak obtained from a plot of spectral density 
using the NCA approximation\cite{Brunner:97}.  
While in the spinless case the atomic level crosses the Fermi
level when it approaches the surface, in the degenerate case $N > 1$
the level sits just above it for a wide range of distances and eventually
evolves into a Kondo resonance during the close encounter.

A simple estimate, within the independent particle picture, of 
the freezing distance $z_{fr}$ at which the charge transfer rate becomes
small gives for $v = 0.01$ a.u. the value $z_{fr} \approx 8$ a.u. 
This means that for perpendicular velocities $u_f \leq 0.01$  
the empty orbital regime is always inside the freezing distance. 
This fact is corroborated in Fig. \ref{fig1} as the charge fractions
for $N > 1$ converge
to those of the independent particle picture at low enough perpendicular 
velocities and at zero temperature.  However, for increasing surface 
temperature, as explained above, the renormalized level becomes populated.   
Analysis of the properties of this level via the spectral 
density led SNL\cite{Shao:96} to conclude
that the integrated charge under the Kondo peak actually freezes in the 
mixed valent regime, that is closer to the surface than what an independent 
particle picture suggests.  This hypothesis explains why the anomalous 
temperature dependence persists even at very low velocities.
Upon increasing the value of $N$ from 2 to 4 or 8, the energy renormalization
increases further and the renormalized level lies  
above the Fermi level even for distances as close as $z = 3$ to $4$ a.u. 

\begin{figure}
\center
\epsfxsize = 10cm \epsfbox{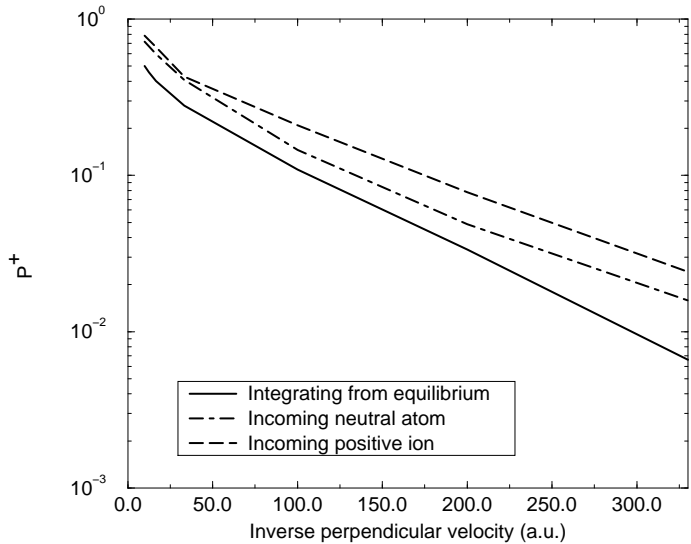}
\caption{
\label{fig6}
Different initial conditions: positive ion or neutral states starting
from far away, with $u_i = u_f$, and the initial equilibrium ground state
starting from
$z_0 = 3$ a.u. The temperature $T = 0$ K, $N = 4$, and we have taken the 
$U \rightarrow \infty$ limit.}
\end{figure}

In light of these results, it is interesting to reconsider how the Kondo 
effect might alter the phenomenon
of loss of memory, the experimentally observed fact that the outgoing charge 
probability distribution of simple atoms such as alkalis and halogens does not
depend on the incoming initial charge state.
A stringent test of loss of memory is to study different initial
conditions for the incoming ion:  start the atom far away from the surface
in a neutral and in a positive ion state.  In Fig. \ref{fig6} we
see that both the neutral and positive yields
are greater than the equilibrium ones.  This breakdown of loss of
memory for the case of the alkaline-earth atoms can be explained
from the fact that the incoming atom creates particle-hole
excitations in the metal which rise the temperature of the metal.
In other words, the surface is heated locally by the formation of particle-hole
pairs during the atom-surface interaction.  Heating is greater
in the case of the incoming positive ion as it induces a larger number of 
particle-hole pairs than an incoming neutral atom, because in both cases
the atom emerges largely neutral.  We find that an 
atom starting from equilibrium at $z_0 = 3$ a.u. with 
the surface at temperature $T = 300$ K gives similar final fractions as 
those obtained for the positive ion scattered off of a zero-temperature surface.

\begin{figure}
\center
\epsfxsize = 10.0cm \epsfbox{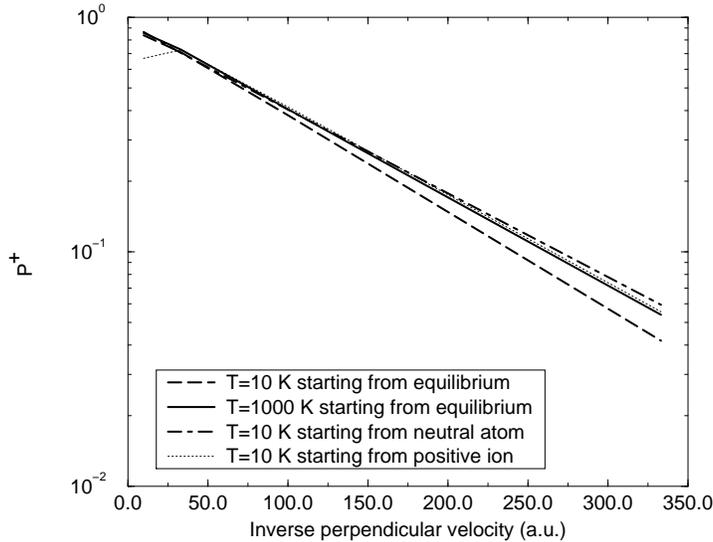}
\caption{
\label{fig7}
Temperature dependence with different initial conditions for the case of an 
alkali atom with the same parameters as used in Fig. 6.} 
\end{figure}

In contrast to the alkaline-earth atoms, we do not expect strong temperature
dependence in the case of alkali atoms such as Li or Na.  
Close to the surface the atom is in the
empty orbital regime, there is no unpaired spin to be screened, and there 
is no renormalization of the atomic level 
energy.  Further out from the surface the local moment regime can occur, but 
the widths are too small for significant charge transfer to occur and the
final fractions do not reflect any temperature dependence. 
Indeed, our calculations verify this picture, as shown in Fig. \ref{fig7}. 
Here the alkali atom has the same parameters
as the alkaline-earth in the above calculations, with
$N = 4$ in the $U \rightarrow \infty$ limit.  Only the nature of the atomic
states has changed.
It is reassuring to see that loss of memory is recovered in the alkali
case and there is a negligible temperature dependence in the final occupancies.  
\subsection{Finite U: Ca atoms bombarding Cu surfaces. }

It was first suggested by SNL\cite{Shao:96} that
alkaline-earth atoms scattered off of noble metal surfaces are good
experimental candidates to exhibit the anomalous temperature
dependence.  Ionization energies of alkaline-earth atoms
are in the range of $4$ to $9$ eV, and work functions of 
noble metal surfaces range between $4$ and $5$ eV. 
The position of the atomic level with respect to the Fermi energy 
should not be greater than about $2$ eV; otherwise the yield of $AE^+$ ions
is too small to be detected experimentally at the small velocities required.
To be specific, we analyze in some detail the problem of calcium  
atoms scattering off copper surfaces. The experimental measured ionization
energy of Ca is $6.1$ eV. The work function of the Cu(001) surface is 
$W = 4.59$ eV.  We now utilize a more realistic three parameter form for the
widths to account for saturation at the chemisorption distance:
\begin{equation}
{\Delta_a(z)} = {{\Delta_0}\over{[{e^{4 \alpha z}}
{}~ +~ {{({\Delta_0}/{\Delta_{sat}})}^4~ -~ 1~ ]}^{1/4}}}\ .
\label{widthfit}
\end{equation}
In all calculations that follow
we take $\Delta_0 = 0.75$ a.u., $\Delta_{sat} = 0.15$ a.u.
and $\alpha = 0.65$, both for the Ca$^0(4s^2)$ and for the Ca$^+(4s^1)$ 
levels.  These parameters represent a ``best guess'' but it 
should be possible to obtain the parameters from 
first-principle, independent-particle calculations\cite{Nordlander:88,More}.

\begin{figure}
\center
\epsfxsize = 10.0cm \epsfbox{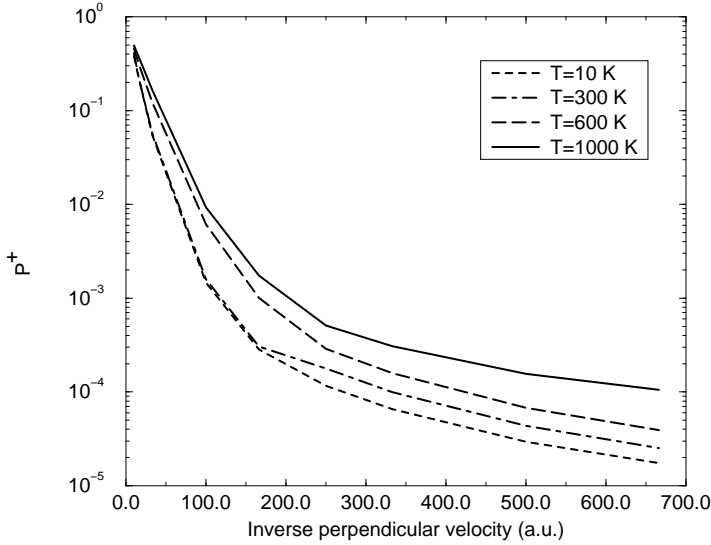}
\caption{
\label{fig8}
Final fractions versus inverse of the velocity for
a Ca atom scattering from a Cu surface.  The atom starts from the equilibrium
ground state at $z_0 = 3$ a.u.}
\end{figure}

We switch $N$ to its physical value, $N = 2$, to model the actual
spin degeneracy of the 4s level.  The value of the $U$ is given
by the difference between the first and second ionization
levels.  The energy required to remove an electron from
the Ca$^+$ 4s orbital is $11.9$ eV, so $U = 5.8$ eV. 
We allow the level to vary in accord with the saturated form given
in Sec. \ref{sec:theory}, namely Eq.(\ref{alkearthshift}).
We take the saturation of the first ionization energy of Ca 
to be $V_{maxI} = 3$ eV.
This choice is justified by recent DF-LCAO calculations which
give the total energies of protons and
He$^{+}$ ions on Al metal\cite{Merino:96,More}. 
As discussed below, the qualitative behavior
of our results do not depend on this choice.
We also choose $V_{maxII} = 3$ eV so that the second ionization energy of
the Ca atom crosses the Fermi energy only for $z < 3$ a.u.      
Although this choice may overestimate the production of Ca$^{++}$
close to the surface, it is a conservative
condition to test the temperature dependence
found above in the $U \rightarrow \infty$ limit.
Recent trajectory simulations of Na atoms scattered from Cu 
surfaces in the hyperthermal energy range\cite{DiRubio:96} suggest a
turning point around $z_0 = 3$ a.u. is reasonable.
In Fig. \ref{fig8} we plot the final fractions 
as a function of the inverse of the perpendicular velocity.  
We have extended the calculations to perpendicular velocities as low as 
$u_f = 0.0015$ a.u.  We see that although the temperature dependence
is somewhat reduced in comparison to the $U \rightarrow \infty$, 
the anomaly remains.  The positive ion yield flattens out at low velocities,
instead of decreasing rapidly, presumably because activity in the 
Ca$^{++}$ channel enhances the Ca$^+$ fraction.  This feature should be
tested experimentally as it indirectly shows the role played by the virtual
Ca$^{++}$ state.  We have also tried doubling the total 
widths of the double hole-occupied state. 
The same temperature dependence is recovered. 

\begin{figure}
\center
\epsfxsize = 10.0cm \epsfbox{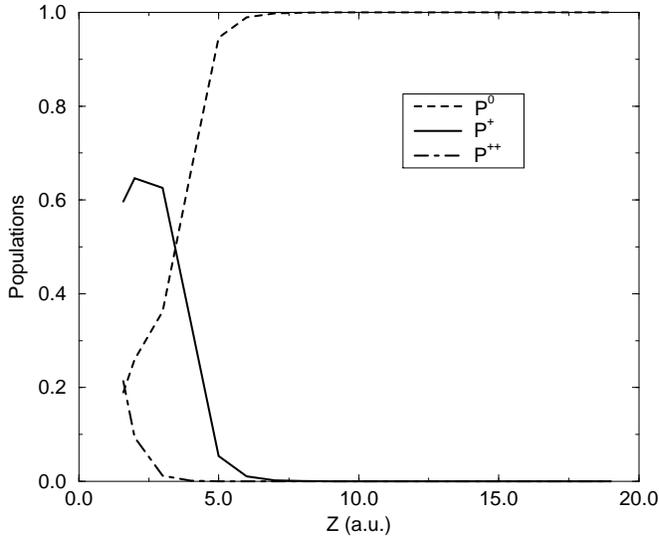}
\caption{
\label{fig9}
Evolution of the different charge fractions along
the trajectory of a Ca atom starting from equilibrium at $z_0 = 3$ a.u. 
with surface temperature set to $T = 10$ K } 
\end{figure}

\begin{figure}
\center
\epsfxsize = 10.0cm \epsfbox{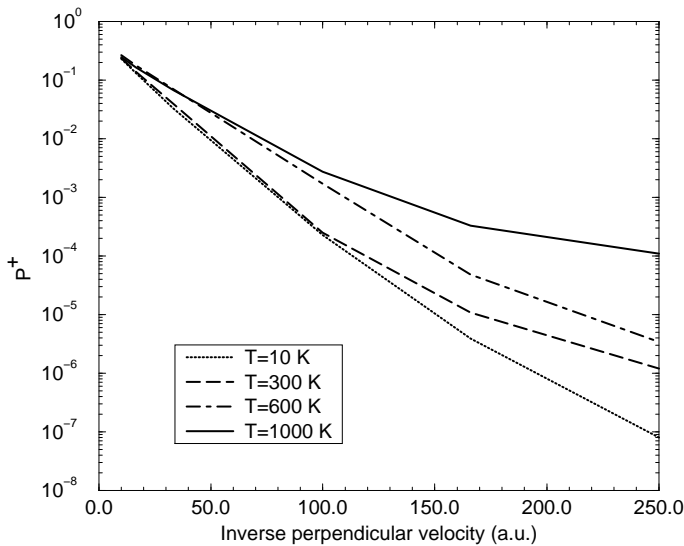}
\caption{
\label{fig10}
Final fractions for the same parameters as in Fig. 8
in the $U \rightarrow \infty $ limit.}
\end{figure}

In Fig. \ref{fig9} we plot the evolution of the charge
fractions along the trajectory of the ion for 
a fixed perpendicular velocity of $u_f = 0.003$ a.u. 
From this figure we see that the Ca$^{++}$ 
recombines into Ca$^+$ quickly due to the three times larger image shift in 
the second ionization energy as a function of distance,
$3e^2 / 4 (z-z_{im})$, compared to the 
first ionization level, which varies as $e^2 / 4 (z-z_{im})$.  
This effect pushes the recombination of Ca$^+$ ions into Ca$^0$ out
to further distances from the surface where the widths are not sufficiently
large for complete neutralization of the positive fractions.  
This explains the enhancement in the probability of Ca$^+$ ions seen 
in Fig. \ref{fig8} for perpendicular velocities as small as $u_f = 0.0015$ a.u. 
at $T = 10$ K.
Finally we plot in Fig. \ref{fig10} 
results for a Ca atom with the same parameters as those used previously
in the limit $U \rightarrow \infty$.  The Ca$^+$ populations decay
more quickly for $T = 10$ K in the $U \rightarrow \infty$ limit
than in the U-finite case.  However, the $T = 1000$ K
plot in Fig. \ref{fig10} resembles the $U = 5.8$ eV case at zero temperature.
For finite-$U$ Haldane's scaling approach can be applied by 
replacing the half-bandwidth $D$ by $U$ in the renormalization 
group formulas\cite{Krishna-murthy:80}.
For $U \approx D$ the position of the renormalized level and the
scaling invariant do not change much from the
$U \rightarrow \infty$ case for distances $z > 5$ a.u. where
perturbative scaling is justified.

Although an analytical expression for the Kondo temperature versus $U$
has been obtained \cite{Schiller:93} it is only valid to order O($1$) 
in the $1/N$ expansion.  We do not expect it to give
a realistic estimate of the Kondo scale for $N = 2$.   
Quantum Monte Carlo calculations of spectral densities
for the symmetric Anderson model suggest an increase
in the Kondo temperature as the value of U is reduced\cite{Gubernatis:93}.
In the asymmetric case, RG calculations have been carried out which give 
phenomenological expressions for the Kondo temperature.
Unfortunately, the validity of these expressions is questionable
for parameters appropriate to the atom-surface scattering problem.
  
In the above results we have only included 
the lowest energy Ca states:  Ca$^0(4s^2)$, Ca$^{+}(4s)$ and Ca$^{++}(3p^6)$. 
We now consider the effect of extending the atomic configurations to
include the excited Ca$^{+}(3p^6 3d)$
state in the calculation.  Its total energy is $7.8$ eV with respect to 
the neutral Ca$^0(4s^2)$ and therefore its ionization energy is $3.2$ eV
measured with respect to the Fermi energy of a clean Cu(001) surface.  
We increase the decay exponent for this $3d$-state to $\alpha = 1.00$ a.u.
and keep the prefactor $\Delta_0$ and the saturation $\Delta_{sat}$ the same.
Final fractions yield the same qualitative temperature dependence found above.
The insensitivity of our result to the inclusion of this state 
suggests that the Kondo effect is robust.

\section{Conclusions}
\label{sec:conclusions}

We have shown that the Kondo screening cloud which surrounds an unpaired 
atomic spin reveals itself in an enhanced neutralization probability at low
surface temperature.  To demonstrate this we employed the dynamical $1/N$ 
expansion, extended to non-zero temperatures.  The accuracy of the $1/N$ 
expansion was improved in comparison to previous work 
with the addition of a new sector at order O($1/N^2$).
We calculated the final charge fractions of alkaline-earth atoms 
for different temperatures in the $U \rightarrow \infty$ limit, and found
qualitative agreement with the NCA approach.  
We also analyzed the size of the temperature dependence for the case of   
degeneracies $N = 1$, 2, 4, and 8.  As expected, the anomalous temperature
dependence is enhanced with increasing degeneracy of the atomic level.
Furthermore, we performed a stringent test of the accuracy
of the $1/N$ expansion by examining the spinless $N = 1$ case. 
The Kondo effect disappears in this case, and also for alkali atoms, 
as it should. 

A simple perturbative RG scaling argument explains
the temperature dependence found in the degenerate cases.  The bare atomic
level is renormalized to just above the Fermi
energy of the metal so that it can be thermally populated. 
Finally, we examined finite physical values of the Coulomb repulsion $U$ and
excited states appropriate for real Ca atoms.
Even though virtual Ca$^{++}$ ions are produced 
close to the surface, they rapidly recombine into Ca$^+$ as the atom
leaves the surface.  Anomalously strong temperature dependence remains, 
although absolute yields differ somewhat from those obtained
in the $U \rightarrow \infty$ limit.

Experimental detection of this effect may be possible, but
several conditions must be satisfied.  First, kinetic energy deposited in
the lattice during the atom-surface impact must remain largely 
decoupled from the electronic degrees of freedom, 
otherwise the electrons in the surface will be heated and mask
the Kondo effect.  Furthermore, both the parallel and perpendicular
components of the velocity must be kept small to avoid
smearing the Fermi surface in the reference frame of the atom.  
Thus $k_F |\vec{u_f}| \ll k_B T$; for $T = 1000$K this
translates to $|\vec{u_f}| < 0.004$ a.u. in the case of a Cu(001) 
surface.  As final fractions smaller than $0.1\%$ are difficult
to detect experimentally, it is important to choose the surface workfunction 
carefully.  We considered the specific case of Ca hitting Cu(001), but 
yields can be increased by using different metallic surfaces with larger
workfunctions.  Finally,
local heating of the surface in the form of particle-hole excitations 
produced as a by-product of charge transfer between the 
atom and the surface will mask the Kondo effect.  Heating the 
surface, rather than cooling it, appears to be the best way to probe the
anomalous temperature dependence, the cleanest signature of the Kondo effect.

{\bf Acknowledgements}
We thank Andone Lavery, Chad Sosolik, Eric Dahl, Barbara
Cooper, Peter Nordlander and David Langreth for helpful discussions.
Computational work in support of this research was performed on Cray computers 
at Brown University's Theoretical Physics Computing Facility. 
J. Merino was supported by a NATO postdoctoral 
fellowship and JBM was supported in part by NSF Grant DMR-9313856.

\end{document}